\documentclass[twocolumn,letterpaper,amsmath,amssymb,floatfix,aps,superscriptaddress]{revtex4}

\usepackage{hyperref}
\usepackage{verbatim}
\usepackage{graphicx}
\usepackage{bm} 
\usepackage{ifpdf} 
\usepackage{dcolumn}  
\usepackage{epsfig}
\usepackage{color}

\def\ul#1#2{\textstyle{\frac{#1}{#2}}}

\newcommand {\vct}[1] {\mathbf {#1}}

\begin{document}
\setlength\arraycolsep{2pt}

\title{Electromagnetic fluctuation-induced interactions in randomly charged slabs}

\author{Vahid Rezvani}
\affiliation{Department of Physics, University of Isfahan, Isfahan
81746, Iran}

\author{Jalal Sarabadani}
\thanks{Present address: Max Planck Institute for Polymer Research, Ackermannweg 10, D-55128 Mainz, Germany}
\affiliation{Department of Physics, Institute for Advanced Studies in Basic Sciences (IASBS), Zanjan 45137-66731, Iran}
\affiliation{Department of Physics, University of Isfahan, Isfahan 81746, Iran}

\author{Ali Naji}
\thanks{Corresponding author -- email: \texttt{a.naji@ipm.ir}}
\affiliation{School of Physics, Institute for Research in Fundamental Sciences (IPM), Tehran 19395-5531, Iran}
\affiliation{Department of Applied Mathematics and Theoretical Physics, Centre for Mathematical Sciences, University of Cambridge, Cambridge CB3
0WA, United Kingdom}

\author{Rudolf Podgornik}
\affiliation{Department of Theoretical Physics, J. Stefan Institute, SI-1000 Ljubljana, Slovenia} 
\affiliation{Department of Physics, Faculty
of Mathematics and Physics, University of Ljubljana, SI-1000 Ljubljana, Slovenia}
\affiliation{Department of Physics, University of Massachusetts, Amherst, MA 01003, USA}

\begin{abstract}
Randomly charged net-neutral dielectric slabs are shown to interact across a featureless dielectric continuum  
with long-range electrostatic
 forces that scale with the statistical variance of their quenched random charge distribution and inversely with the
distance between their bounding surfaces. By accounting for the whole spectrum of electromagnetic 
field fluctuations, we show that this long-range disorder-generated  interaction extends well into the retarded
regime where higher-order Matsubara frequencies contribute significantly. This occurs even for highly clean samples with
only a trace amount of charge disorder and shows that disorder effects can be important down to the nano scale. 
As a result, the previously predicted non-monotonic behavior for the total force between dissimilar 
slabs as a function of their separation distance is substantially modified 
by higher-order contributions, and in almost all cases of interest, we find that the equilibrium inter-surface separation 
is shifted to  substantially larger values compared to predictions based solely on the zero-frequency component. This suggests that the ensuing
non-monotonic interaction is more easily amenable to experimental detection. 
The presence of charge disorder in the intervening dielectric medium between the two
slabs  is shown to lead to an additional force that can be repulsive or attractive depending
on the system parameters and can, for instance,
wash out the non-monotonic behavior of the total force when the intervening slab contains a sufficiently large amount of disorder charges.
\end{abstract}
\maketitle

\section{Introduction} 

Patterned and heterogeneously charged materials, in particular if the heterogeneity is disorder induced, have received much attention in recent years in a number of 
different research areas. For instance,  randomly charged polyelectrolytes \cite{rand_polyelec} and patchy colloids \cite{patchy_colloids} show distinct
collective and thermodynamic properties than ordinary colloids and charged homopolymers \cite{DLVO,Doi}. 
Proteins for instance represents a prime example
of biological molecules exhibiting heterogeneous and  highly disordered charge distributions. 
The high specificity and selectivity of protein-protein interactions is one of the fundamental problems
of molecular biology and requires an understanding of the interaction between randomly patterned surfaces \cite{protein}. Another 
example which has been in the focus of recent experimental investigations is the problem of interaction between surfactant-coated
surfaces which exhibit unusually strong and long-range attractive forces \cite{surf}, shown to stem directly from the presence of
quenched random domains (patches) of positive and negative charges on these surfaces \cite{surf_new}. 

In fact, most solid surfaces exhibit heterogeneous charge distributions that can be highly disordered as revealed by recent Kelvin force 
microscopy measurements \cite{science11}. Such random charges may result from the surface adsorption of charged 
contaminants and/or impurities, while even clean polycrystalline samples display patchy surface 
potentials \cite{barrett,speake}. The patchiness of the surface potential
is believed to lead to significantly large effects in the experiments aimed at measuring the Casimir-van der 
Waals (vdW) interactions between
solid surfaces in vacuum. Indeed, recent ultra-high sensitivity measurements have shown the presence of an  
``anomalously" long-range interaction which can easily mask the Casimir-vdW force at sufficiently large separations \cite{kim1,kim2,kim3,kim4,tang}. 

In a series of theoretical papers \cite{epl2006,prl2010,jcp2010,pre2011,epje2012}, 
the effects of quenched monopolar charge disorder in the bulk or surface
of dielectric slabs were investigated. It was shown that even a small amount of quenched random charges can lead to strong long-range
interactions between dielectric slabs. These interactions were shown to result directly from the interplay between the electrostatic interactions 
generated by the presence of dielectric discontinuities (the so-called image charge effects) and the quenched statistics of the random
charges. It is thus remarkable to note that such forces exist even for dielectrics which are {\em electroneutral} on the average but carry a disordered
charge component. In this case, net Coulomb forces are obviously absent and thus the disordered-induced forces directly compete 
with the Casimir-vdW forces. While the latter dominates at small separations, 
the former becomes substantially large and wins at large separations. The previous calculations were however performed only within the classical
regime, where only the zero Matsubara frequency contributes to the Casimir-vdW force  \cite{prl2010,jcp2010}. Strictly speaking, this approximation
would be valid above the thermal wavelength (around 7 microns at room temperature) \cite{vdWgeneral}, 
although the contribution from higher-order Matsubara frequencies would in fact dominate
 at much smaller separations depending on the dielectric properties of the materials 
(e.g., below about 1 micron in vacuum and 100~nm in a polar medium such as water \cite{ninham}). 
The zero-frequency results would be relevant for the large-distance regime where the 
above-mentioned anomalous force is observed \cite{kim1,kim2,kim3,kim4,tang}. However, at 
sub-micron separations, it would be necessary to examine the quantum effects 
from the higher-order Matsubara modes of the electromagnetic field fluctuations. 

In the present work, we shall thus set out to investigate in detail the interaction between two randomly charged 
net-neutral dielectric slabs by accounting for the full spectrum of 
electromagnetic field fluctuations in the following two cases: i) the slabs interact across a disorder-free 
dielectric continuum, considering both similar as well as dissimilar slabs, and ii) 
the slabs are separated in general by a dielectric layer which itself may contain random quenched charges. 
In both cases, the results can be compared directly against those reported previously \cite{prl2010,jcp2010}; hence, 
we can determine the effects of the inclusion of higher-order Matsubara frequencies, which will be computed 
via the Lifshitz formalism \cite{vdWgeneral,jalal}, as well as the quenched disorder 
charges in the intervening medium. 

These results thus generalize the analysis of the disorder effects to all ranges of separation down to 
the nano scale (as long as the continuum dielectric model employed
within the Lifshitz formalism remains valid). We can then draw conclusions regarding the crossover between 
different scaling regimes for the interaction between slabs, which were missing from a zero-frequency 
analysis \cite{prl2010,jcp2010}.  In particular, we show that the characteristic 
$\sim D^{-1}$ decay \cite{prl2010} of the total force with the distance $D$ between two randomly 
charged but otherwise (dielectrically) identical semi-infinite slabs  
sets in well within the retarded regime (around, e.g., 50-500~nm). Therefore,  it is found that  the interaction 
crosses over to this $\sim D^{-1}$ disorder-induced  behavior from the standard (retarded) 
$\sim D^{-4}$ Casimir-vdW behavior rather than from the classical zero-frequency $\sim D^{-3}$ behavior. 
It turns out that even for  highly clean samples (with disorder charge
densities down to $10^{-9}\, {\mathrm{nm}}^{-3}$), the magnitude of the disorder-induced force is substantial
if compared with the Casimir-vdW force. 

For dielectrically dissimilar slabs, we show that the non-monotonic behavior \cite{jcp2010} of the total force as a function
of distance persists when higher-order Matsubara frequencies are included. However  in almost all cases, the equilibrium 
separation defined through the zero of total force (which can represent either a stable or unstable free 
energy extremum) is shifted to separations that can be substantially larger than those predicted
within the zero-frequency theory \cite{jcp2010}. This is an important 
consequence of our analysis and suggests that the non-monotonic features of the interaction force between 
dielectric slabs could be easily amenable to experimental 
measurements in this regime \cite{tang}. Such non-monotonic interaction profiles have received a lot of 
attention in the context of the Casimir effect in recent years and may arise  in the case of metamaterials \cite{metamat}
and/or other exotic materials such as topological insulators \cite{topol}, as well as in certain non-trivial
geometries \cite{geom_casimir,sernelius}.  In our analysis the  $\sim D^{-1}$  behavior of the interaction force for identical 
slabs and the  non-monotonic force profile for dissimilar slabs represent characteristic fingerprints of the 
charge disorder and can thus be useful in assessing 
whether the experimentally observed interactions in ordinary dielectrics can be interpreted  in terms of disorder effects.  

The organization of the paper is as follows:
In Section \ref{sec:model}, we introduce our model and the details of 
the formalism employed in our analysis. The results for two semi-infinite slabs interacting across
vacuum or a disorder-free dielectric layer are discussed in Section \ref{sec:no_dis} and those for the case 
where the slabs interacting across a dielectric layer which itself contains disordered charges is discussed in 
Section \ref{sec:dis}. We conclude our study in Section \ref{sec:discussion}.

\section{Model and Formalism}
\label{sec:model}

\begin{figure}[t]
\includegraphics[angle=0,width=7cm]{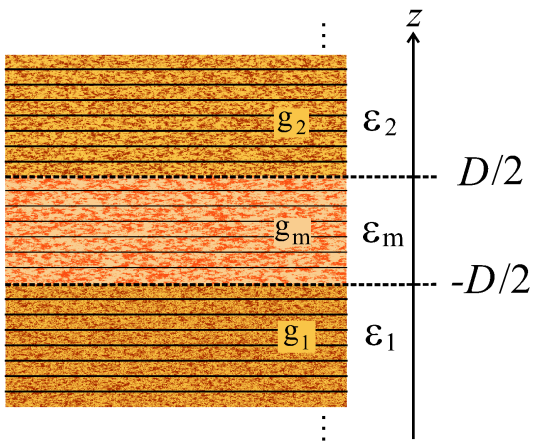}
\caption{ (Color
online) We consider two semi-infinite net-neutral regions with dielectric response functions 
$\epsilon_{1}(\omega)$ and $ \epsilon_{2}(\omega)$ with the static dielectric constants, 
$\varepsilon_1$ and
$\varepsilon_2$, respectively, and monopolar charge disorder distributions (shown
schematically by small light and dark patches) 
described with variances $g_1$ and $g_2$. The intervening slab may be either vacuum or in general a net-neutral dielectric
material of thickness $D$ and dielectric response function $\epsilon_{m}(\omega)$ with the static dielectric 
constant of $\varepsilon_m$ and charge disorder of variance $g_m$. }
\label{fig:schematic}
\end{figure}

We consider a plane-parallel three-slab system consisting of two semi-infinite regions of dielectric response functions 
$\epsilon_{1}(\omega)$ and $ \epsilon_{2}(\omega)$ and an intervening slab of thickness $D$ and dielectric response function
$\epsilon_{m}(\omega)$ (see Fig. \ref{fig:schematic}). All three slabs are assumed to carry a disordered
{\em monopolar charge} distribution, $\rho(\vct r)$  \cite{note_gs}. The disordered charge distribution is taken to 
have a zero mean value $\langle \! \langle  \rho(\vct r)  \rangle   \! \rangle = 0$, which ensures that the slabs are {\em net-neutral}, 
and a two-point correlation function 
\begin{equation}
\langle \! \langle  \rho(\vct r) \rho(\vct r ') \rangle   \!
\rangle = {\mathcal G}({\boldsymbol \varrho} - {\boldsymbol \varrho}'; z) \delta (z- z'),
\label{charge_correlation}
\end{equation}
where  $\langle \! \langle  \cdots  \rangle   \! \rangle$ denotes
the average over all realizations of the charge disorder
distribution, $\rho(\vct r)$. Here, ${\boldsymbol \varrho} = (x, y)$ denotes
the lateral directions in the plane of the slab perpendicular to the
$z$ axis where the bounding surfaces are taken to be located at $z=\pm D/2$.  
The above form of the correlation function thus implies no spatial correlations in $z$ 
direction and can be thus applicable in general to layered materials. In lateral directions, we have a statistically invariant correlation function whose 
specific form may depend on $z$, i.e.
\begin{equation}
{\mathcal G}({\boldsymbol \varrho} - {\boldsymbol \varrho}'; z)  =  g(z)c({\boldsymbol \varrho} - {\boldsymbol \varrho}'; z), 
\label{eq:bs_decomp}
 \end{equation}
where 
\begin{eqnarray}
c({\mathbf x}; z) &=&  \left\{
\begin{array}{ll}
     c_{1}({\mathbf x})  & \quad  z<-D/2,\\
          c_{m}({\mathbf x})      & \quad  |z|<D/2,\\
     c_{2}({\mathbf x})  & \quad  z>D/2,
\end{array}
\right.
\label{eq:c_b}
\end{eqnarray}
and we shall further assume that the disorder variance $g(z)$, which gives the density of random quenched charges in the bulk of the
slabs \cite{prl2010}, is given by
\begin{eqnarray}
g(z) &=&  \left\{
\begin{array}{ll}
     g_{1}e_0^2  & \quad  z<-D/2,\\
     g_{m}e_0^2           & \quad  |z|<D/2,\\
     g_{2}e_0^2  & \quad  z>D/2.
\end{array}
 \label{eq:g_b}
\right.
\end{eqnarray}

We shall not deal with effects due to  disorder in the dielectric response of the interacting 
media  \cite{randomdean1,randomdean2}, which presents an additional source of disorder meriting further study and
focus only on quenched disorder  (see Refs. \cite{epl2006,prl2010,jcp2010,mama,pre2005} for the cases studied with annealed or partially annealed disorder,
or with mobile ions on or in between the randomly charged surfaces on the zero-frequency level). The quenched model is an idealization of the real nature of random charge distributions that can
in general exhibit dynamical behavior, but these effects are expected to be small since ion relaxation processes are 
extremely slow on the scale of the Matsubara frequencies. 

We base our analytical calculations on no other assumption regarding the lateral correlation function, so the 
expressions in what follows may be applied straightforwardly to some rather general cases, such as
disorder distributions with a ``patchy" structure characterized 
by a lateral correlation function decaying over a finite correlation length \cite{jcp2010}. Although, for the sake of brevity, 
we restrict our final discussion in this paper to the case where the disorder distribution is statistically homogeneous and uncorrelated, 
$c({\boldsymbol \varrho} - {\boldsymbol \varrho}'; z) = \delta({\boldsymbol \varrho} - {\boldsymbol \varrho}')$, thus 
\begin{equation}
\langle \! \langle  \rho(\vct r) \rho(\vct r ') \rangle   \!
\rangle = g(z) \delta (\vct r - \vct r'). 
\label{eq:twopoint_uncorr}
\end{equation}

In our previous works, we  derived the partition function of the system defined above for the case where 
the intervening medium is free from any kind 
of charge disorder and the electromagnetic field fluctuations are taken into account only on the 
zero-frequency level \cite{prl2010,jcp2010}. The latter would be a valid approximation only at 
sufficiently large separation distances, $D$, or sufficiently high temperatures, $T$. In the present work, we shall account for 
all higher-order Matsubara modes of the electromagnetic field fluctuations, which become increasingly 
important at small separations  down to the nano scale. It is easy to see that when the disorder 
is perfectly quenched, as indeed we assume here, these charge sources only couple to the zero-frequency mode and thus do not
mix with the higher-order frequency modes of the electromagnetic field fluctuations. We do not delve further 
into the details of the derivation of the free energy of the quenched
system, which can be written, after averaging over various realizations of the charge 
disorder  (see Ref. \cite{jcp2010} for details), in an additive form as 
\begin{equation}
{\mathcal F}= {\mathcal F}_{\mathrm{vdW}} + {\mathcal F}_{\mathrm{dis}}.
\label{free_energy_quenched}
\end{equation}
The first term on the right hand side above is the Casimir-vdW interaction free energy, which is obtained in
the Lifshitz form of the surface free energy density as
\begin{eqnarray}
\frac{\beta {\mathcal F}_{\mathrm{vdW}}}{S} &=& \sum_{\bf Q}
{\sum_{n=0}^{\infty}}^\prime  \ln \bigg[ 1 - \Delta^{({\mathrm{TM}})}_{2,m}(\imath \xi_n) \Delta^{({\mathrm{TM}})}_{1,m} (\imath \xi_n) \times
 \nonumber\\ && \times \, e^{-2 D \kappa_{m}(\imath \xi_n)} \bigg]  + ~[({\mathrm{TM}}) \rightarrow ({\mathrm{TE}})],
\label{eq:vdw_Free-Energy}
\end{eqnarray}
where $k_{\mathrm{B}}T=1/\beta$ and $S$ is the surface area of the slabs. The free energy is normalized  in such 
a way  that it tends to zero at infinite separation distance $D$ and TM and TE correspond to transverse 
magnetic and transverse electric modes. In the Lifshitz formula
the $\bf Q$ summation is over the transverse wave vector and the $n$ summation (where
the prime indicates that the $n=0$ term has a weight of $1/2$) is over the imaginary Matsubara frequencies 
\begin{equation}
\xi_n=\frac{2 \pi n k_{\mathrm{B}} T}{\hbar},
\end{equation} 
where $\hbar$ is the Planck constant divided
by $2 \pi$. All the quantities in the bracket depend on $\bf Q$ as well as $\xi_n$.
We have defined
\begin{equation}
\Delta^{({\mathrm{TM}})}_{\alpha,\beta}(\imath \xi_n) = \frac{ \epsilon_\alpha(\imath \xi_n) \kappa_{\beta}(\imath \xi_n) - \epsilon_{\beta}(\imath \xi_n)
\kappa_{\alpha}(\imath \xi_n) }{\epsilon_\alpha(\imath \xi_n) \kappa_{\beta}(\imath \xi_n) + \epsilon_{\beta}(\imath \xi_n) \kappa_{\alpha}(\imath \xi_n)},  
\label{eq:Delta-Two-sheets}
\end{equation}
which quantify the dielectric mismatch across the bounding surfaces between the three different slabs labeled by 
$\alpha, \beta=1, m, 2$. Also $\kappa_\alpha(\imath \xi_n)$ for each electromagnetic field  
mode within the slab $\alpha$  is given by
\begin{equation}
\kappa_{\alpha}^{2} (\imath \xi_n) =  Q^2 + \frac{ \epsilon_\alpha(\imath
\xi_n)  \mu_\alpha(\imath \xi_n) \xi_n^2 }{c^2},
\label{eq:rho-Matsubara-Two-sheets}
\end{equation}
where $c$ is the speed of light {\em in vacuo}, $Q$ is the magnitude of
the transverse wave vector, and $\epsilon_\alpha(\imath \xi_n)$ and
$\mu_\alpha(\imath \xi_n)$ are the dielectric response function and the magnetic
permeability of the corresponding slab at imaginary Matsubara frequencies, respectively.
For the sake of simplicity we assume that for all slabs $\mu_\alpha(\imath \xi_n)=1$. 
Note that $\epsilon(\imath \xi)$ is standardly referred to as the vdW-London dispersion transform of the dielectric function
and follows as \cite{Wooten}
\begin{equation}
\epsilon(\imath\xi)=1+\frac{2}{\pi}\int_{0}^{\infty}\frac{\omega\,  {\mathrm{Im}}[\epsilon(\omega)]}{\omega^{2}+\xi^{2}}\, d\omega, 
\label{eq:KK_transform}
\end{equation}
being in general a real, monotonically decaying function of the imaginary
argument $\xi$ \cite{vdWgeneral,ninham}.

For the TE modes everything remains the same except that in this case
\begin{equation}
\Delta^{({\mathrm{TE}})}_{\alpha,\beta}(\imath \xi_n) = \frac{ \kappa_{\beta}(\imath \xi_n) - 
\kappa_{\alpha}(\imath \xi_n) }{ \kappa_{\beta}(\imath \xi_n) +  \kappa_{\alpha}(\imath \xi_n)}.
\label{eq:DeltaTE-Two-sheets}
\end{equation}

The general contribution from disorder charges follows in an exact form as \cite{jcp2010}
\begin{eqnarray}
\label{eq:F_quenched}
{\mathcal F}_{\mathrm{dis}}  &=& {\ul{1}{2}}  \int\! {\mathrm{d}}{\mathbf r}\,{\mathrm{d}}{\mathbf r}'\, {\mathcal G}({\boldsymbol \varrho} - {\boldsymbol \varrho}'; z)
\delta(z-z') G({\vct r}, {\vct r}') = \nonumber\\
&&\hspace{-1.2cm} = {\ul{1}{2}} \int\! {\mathrm{d}}{\mathbf r}\,{\mathrm{d}}{\mathbf r}'\,g(z) c({\boldsymbol \varrho}  -
{\boldsymbol \varrho}'; z) \delta(z-z')G({\vct r}, {\vct r}').
\end{eqnarray}
where $G({\vct r}, {\vct r}')$ is the zero-frequency (electrostatic) Green's function defined via
\begin{equation}
 \varepsilon_0 \nabla\cdot [\varepsilon(\vct r) \nabla  G(\vct r, \vct r')] = -\delta(\vct r - \vct r'), 
 \label{green1}
\end{equation}
for the {\em zero-frequency} or {\em static} dielectric constant profile defined as
\begin{eqnarray}
\varepsilon({\mathbf  r})  &=&  \left\{
\begin{array}{ll}
     \varepsilon_1\equiv \epsilon_1(0)&   \quad z<-D/2,\\
        \varepsilon_m \equiv \epsilon_m(0)   & \quad   |z|<D/2,\\
      \varepsilon_2 \equiv \epsilon_2(0)  &   \quad  z>D/2.
\end{array}
 \label{eq:epsilon}
\right.
\end{eqnarray}

Equation (\ref{eq:F_quenched}) is valid for any
arbitrary disorder correlation function $ {\mathcal G}({\boldsymbol \varrho} - {\boldsymbol \varrho}'; z) $ and dielectric constant profile $\varepsilon({\vct r})$.
For the particular plane-parallel three-slab model considered in this work, the Green's function  $G({\vct r}, {\vct r}')$ 
can be calculated from standard methods and one finds
\begin{eqnarray}
\frac{ \beta {\mathcal F}_{\mathrm{dis}}} {S}& = &  - l_{\mathrm{B}} 
\int  \! \frac{{\mathrm{d}}Q}{Q}
\frac{e^{-2 Q D}}{1 - \Delta_1 \Delta_2 \, e^{-2 Q D}} \bigg[ 
\frac{\varepsilon_m g_{1} c_{1}(Q)}{ (\varepsilon_1 + \varepsilon_m)^2 }\, 
\Delta_2  + \nonumber\\
&& \hspace{-1.4cm}\!+ \frac{\varepsilon_m g_{2} c_{2}(Q)}{ (\varepsilon_2 + \varepsilon_m)^2   }  \Delta_1  
\!- \!\frac{\varepsilon_1 g_{m} c_{m}(Q)}{ (\varepsilon_1 + \varepsilon_m)^2 }  \Delta_2
\!- \!\frac{\varepsilon_2 g_{m} c_{m}(Q)}{ (\varepsilon_2 + \varepsilon_m)^2 } \Delta_1
\!\bigg],
\label{eq:F_dis}
\end{eqnarray}
for arbitrary separation distance $D$, where 
\begin{equation}
\Delta_i = \frac{\varepsilon_{i}  - \varepsilon_{m}}{\varepsilon_{i}  +
\varepsilon_{m}} \qquad\quad i=1, 2, 
\end{equation}
is the static dielectric jump parameter at each of the bounding surfaces at $z=\pm D/2$, and 
\begin{equation}
l_{\mathrm{B}}= \beta e_0^2 /(4 \pi {\varepsilon_0}) 
\end{equation}
 is the Bjerrum length in vacuum  ($l_{\mathrm{B}}\simeq 56.8 $~nm at room temperature),
and $c_\alpha(Q)$ is the Fourier transform of the correlation function $c_\alpha({\mathbf x})$. 
As noted before, we shall focus here on the particular case with no spatial correlations, see 
Eq. (\ref{eq:twopoint_uncorr}), corresponding to $c_\alpha(Q)=1$ for all three slabs $\alpha=1, m, 2$. 

Note that  ${\mathcal F}_{\mathrm{dis}}$ stems from electrostatic interactions between randomly 
distributed disorder charges in the three slabs. Due to the dielectric discontinuities across the two 
bounding surfaces, each disorder charge is accompanied by an infinite number of electrostatic  ``images", which 
are in fact generated by the (static) polarization of the slabs. 
These  ``image" charges also contribute to the total free energy of the system as they interact among themselves and with the
actual disorder charges. This type of effects are systematically taken into account through the electrostatic
Green's function $G({\vct r}, {\vct r}')$ and are completely included in the above disorder free energy \cite{jcp2010}.  

Our goal is to calculate the effective interaction force $f$, which is mediated  between slabs 1 and 2 by both 
the electromagnetic field fluctuations and the disorder charges 
placed in the intervening slab, or equivalently the effective interaction free energy ${\mathcal F}$ between the 
two bounding surfaces at $z=\pm D/2$, i.e. 
\begin{equation}
f = -\frac{\partial {\mathcal F}}{\partial D} \qquad {\mathrm{with}} \qquad {\mathcal F}(D) =  {\mathcal F}_{\mathrm{vdW}}(D) + {\mathcal F}_{\mathrm{dis}}(D),
\label{eq:tot_f}
\end{equation}
which can thus be calculated in an explicit form from Eqs. (\ref{eq:vdw_Free-Energy}) and (\ref{eq:F_dis}).

\begin{figure*}[t]\begin{center}
	\begin{minipage}[b]{0.375\textwidth}\begin{center}
		\includegraphics[width=\textwidth]{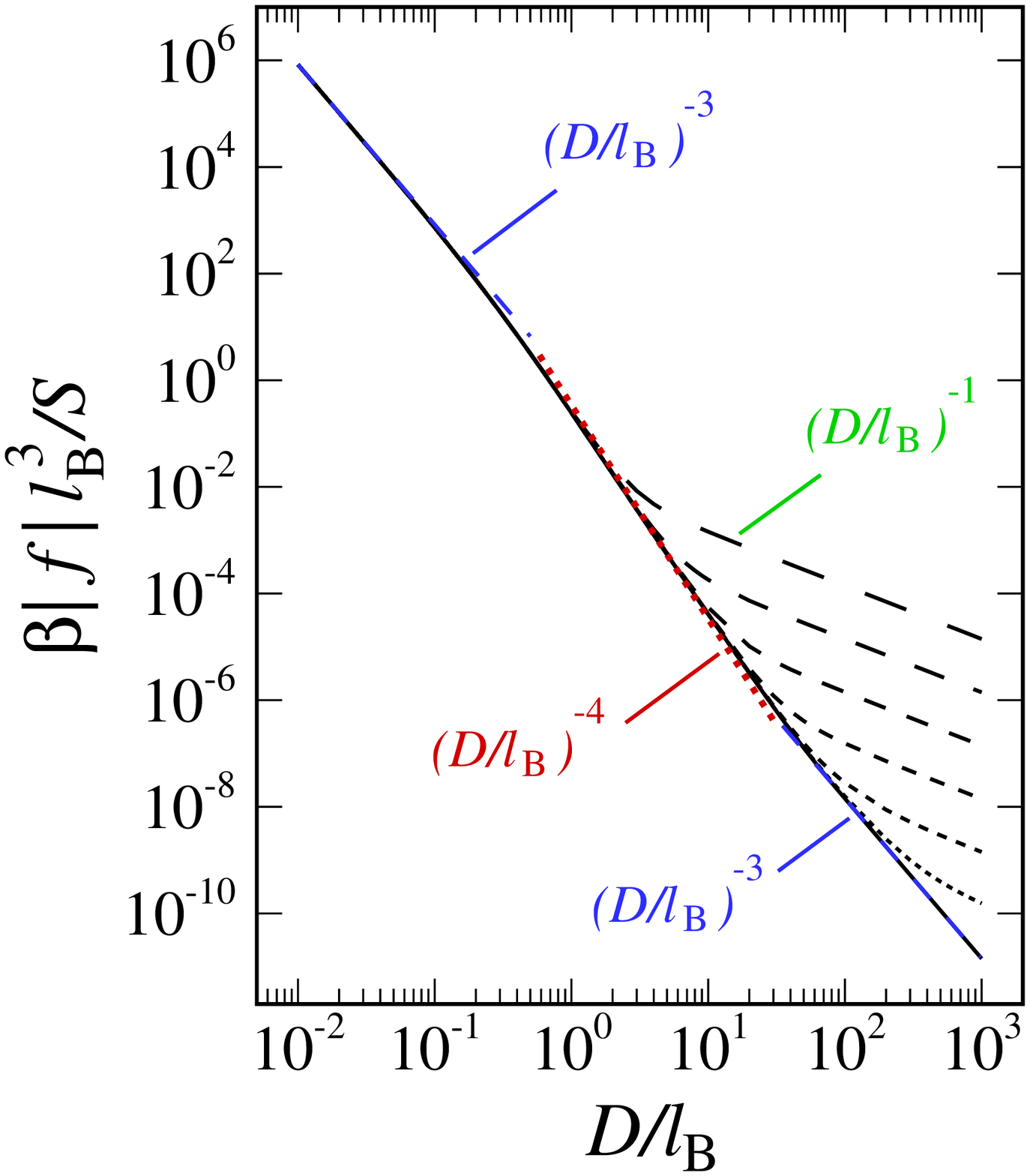} (a)
	\end{center}\end{minipage} \hskip-1.3cm
	\begin{minipage}[b]{0.375\textwidth}\begin{center}
		\includegraphics[width=\textwidth]{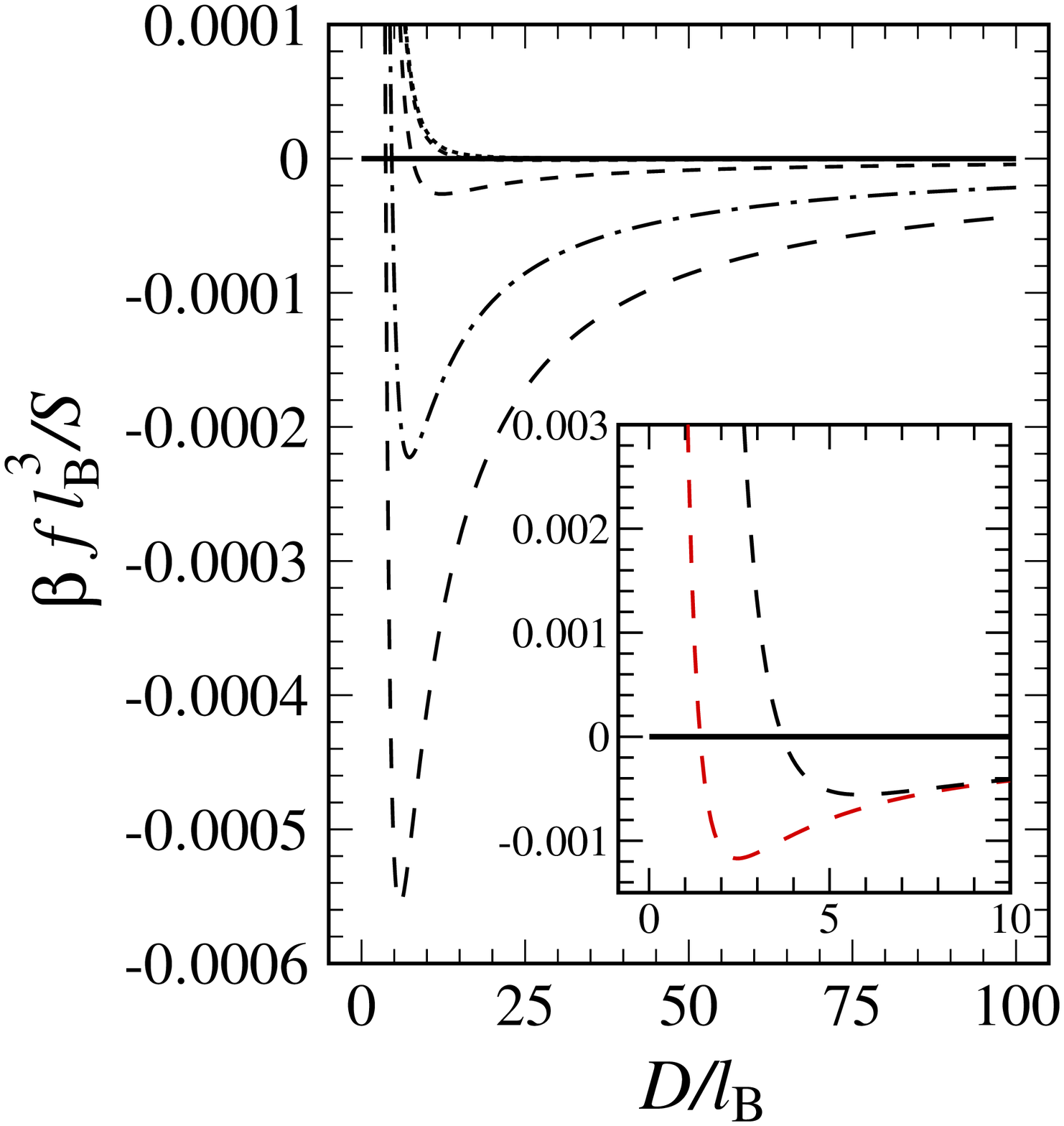} (b)
	\end{center}\end{minipage} \hskip-1.3cm
	\begin{minipage}[b]{0.375\textwidth}\begin{center}
		\includegraphics[width=\textwidth]{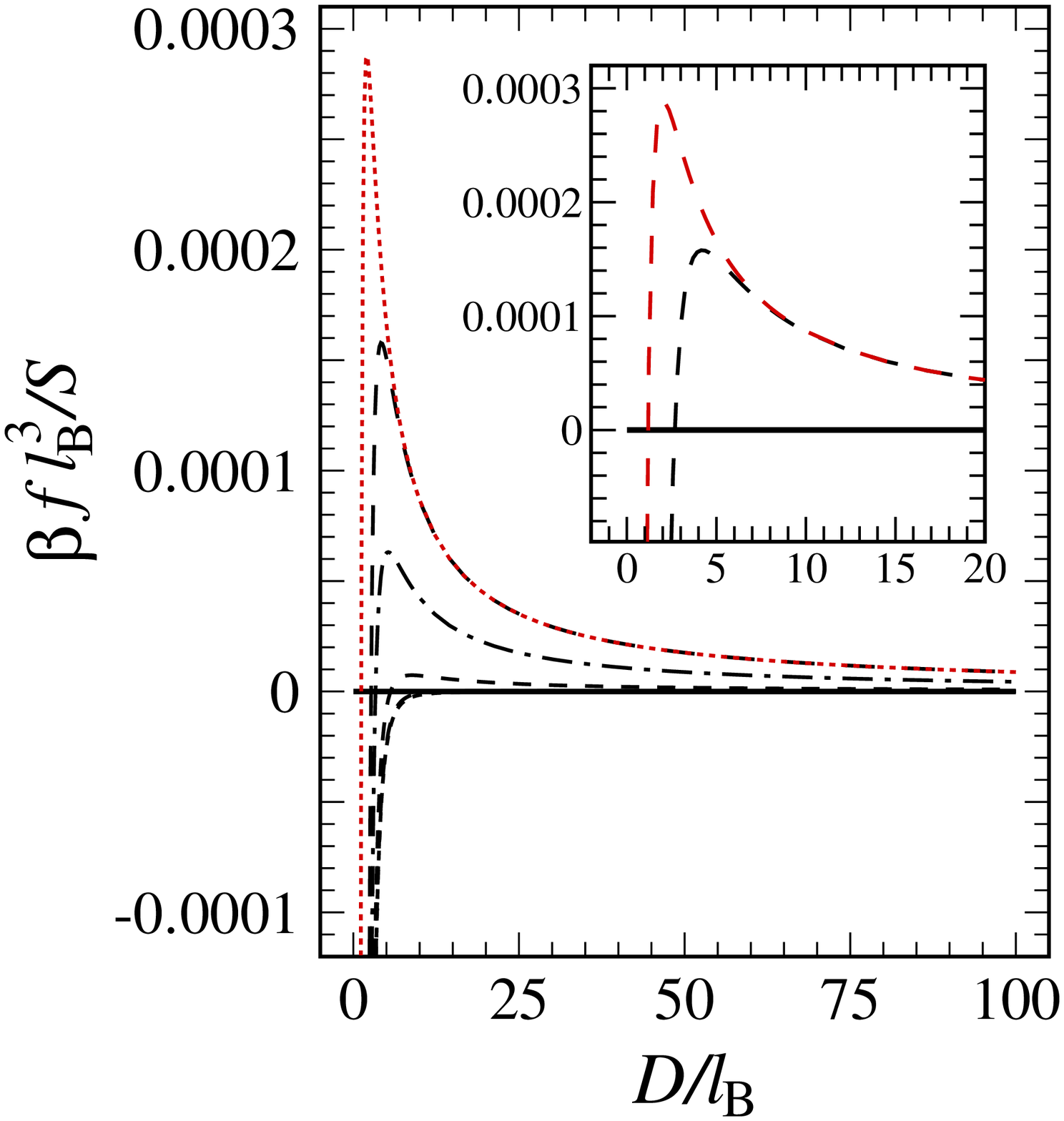} (c)
	\end{center}\end{minipage} \hskip-1.3cm	
\caption{(Color
online)  (a) Magnitude of the rescaled total force, $\beta |f| l_{\mathrm{B}}^3 /S$, between
two identical semi-infinite and net-neutral dielectric slabs in {\em vacuum}
($\varepsilon_m =1$) is plotted on a log-log scale as a function of the rescaled distance, $D/l_{\mathrm{B}}$
(see  Eq. (\ref{eq:tot_f})). The dielectric response 
function is taken according to Eq. (\ref{eq:epsilon-substrate}) with $\varepsilon_1=\varepsilon_2=3.81$, which is appropriate for SiO$_2$ (see the text for other parameters). 
The bulk disorder variance is varied 
in the range $g_1=g_2 = 10^{-6}, 10^{-7}, 10^{-8}, 10^{-9}, 10^{-10}, 10^{-11}\, {\mathrm{nm}}^{-3}$ (from top
to bottom). Solid curve shows the pure Casimir-vdW force obtained from Eq.  (\ref{eq:vdw_Free-Energy}).  
The scaling behavior  of the free energy  in various regimes of separation is shown explicitly.
(b) The rescaled total force, $\beta f l_{\mathrm{B}}^3
/S$ between two dissimilar net-neutral  slabs when the dielectric response functions satisfy the relationship
$\epsilon_1(\imath \xi)<\epsilon_m(\imath \xi)<\epsilon_2(\imath \xi)$ with the static dielectric constant values
$\varepsilon_1=5$, $\varepsilon_m = 10$ and $\varepsilon_2=50$. Here we assume $g_m=0$ and $g_1=g_2 = 10^{-6}, 5\times 10^{-7}, 10^{-7}, 10^{-8}, 10^{-10}\, {\mathrm{nm}}^{-3}$ (from 
bottom to top). Inset shows  a closer view of the region around the minimum for $g_1=g_2 = 10^{-6}\, {\mathrm{nm}}^{-3}$, compared with
the results obtained by including only the zero-frequency contribution (red curve). 
(c) Same as (b) but here we take $\epsilon_1(\imath \xi), \epsilon_2(\imath \xi)<\epsilon_m(\imath \xi)$ with the static dielectric constant values
$\varepsilon_1=15$, $\varepsilon_m = 30$ and $\varepsilon_2=25$. 
}
\label{fig:fig2}
\end{center}
\end{figure*}

\section{Results}

\subsection{Role of higher-order Matsubara frequencies}
\label{sec:no_dis}

In order to bring out the role of higher-order Matsubara frequencies and compare it with our previous zero-frequency results \cite{jcp2010},
we shall first proceed by taking two slabs interacting across vacuum or a disorder-free dielectric medium with 
$g_m=0$. We consider three different cases: two identical slabs with $\epsilon_1(\imath \xi)=\epsilon_2(\imath \xi)$ in vacuum and
two dissimilar slabs with $\epsilon_1(\imath \xi)<\epsilon_m(\imath \xi)<\epsilon_2(\imath \xi)$ and $\epsilon_1(\imath \xi), \epsilon_2(\imath \xi)<\epsilon_m(\imath \xi)$. 
In order to enable a direct comparison between these different cases, we assume a simple model for the vdW-London dispersion transform of the 
dielectric response functions of the slabs as \cite{ninham}
\begin{equation}
\epsilon (\imath \xi)
=1+\frac{C_1\omega_1^2}{\xi^2
+\omega_1^2}+\frac{C_2\omega_2^2}{\xi^2
+\omega_2^2}, \label{eq:epsilon-substrate}
\end{equation}
in all cases. This form for the vdW-London dispersion transform mimics two characteristic relaxation mechanisms in the materials (e.g., 
one due to electronic polarization and the other due to ionic polarization as is the case for SiO$_2$ \cite{Hough}). 
The parameters $C_1$, $C_2$, $\omega_1$ and $\omega_2$ 
are chosen such that the required inequality
relationships between different dielectric response functions at all Matsubara frequencies  as well as the desirable values for the zero-frequency 
dielectric constants are obtained. These values are given in Table \ref{table1} and are not meant to represent any specific material. The qualitative
aspects of our results do not depend on the particular  values for these parameters but on the relationships between dielectric
response functions of the slabs as discussed throughout the text. 
All calculations that follow are done at room temperature $T=300~\mathrm{K}$.

\begin{table}[b!]
	\begin{center}
\begin{tabular}{|c|c|c|}
\hline
 ~ $\varepsilon\equiv \epsilon(0)$ ~&~ $C_1$ ~&~ $C_2$  ~\\
\hline
 ~ 3.81 (SiO$_2$) ~&~ 1.098 ~&~ 1.703 ~\\
 ~ 5 ~&~ 1 ~&~ 3 ~\\
 ~ 10 ~&~ 3 ~&~ 6 ~\\
 ~ 15 ~&~ 5 ~&~ 9 ~ \\ 
 ~ 25 ~&~ 9 ~&~ 15~  \\
 ~ 30 ~&~ 11 ~&~ 18 ~ \\
 ~ 40 ~&~ 14 ~&~ 25 ~ \\
 ~ 50 ~&~ 20 ~&~ 29 ~\\ 
 ~ 60 ~&~ 24 ~&~ 35 ~ \\ 
 ~ 100 ~&~ 34 ~&~ 65~  \\ 
\hline
\end{tabular}
\caption{Parameter values for the coefficients $C_1$ and $C_2$ in 
the  vdW-London dispersion transform of the 
dielectric response function (\ref{eq:epsilon-substrate}) are chosen in such a way that the
static dielectric constants, $\varepsilon\equiv \epsilon(0)$, shown on the left column are reproduced. 
In all cases, we fix the characteristic frequencies as $\omega_1 = 2.033 \times 10^{16}$ rad/s and
$\omega_2 = 1.88 \times 10^{14}$ rad/s.}
\label{table1}
\end{center}
\end{table}

Let us first consider the case of two identical slabs with equal charge disorder densities, $g_1=g_2=g$, in vacuum. 
In this case, we choose the parameters in Eq. (\ref{eq:epsilon-substrate}) appropriate for SiO$_2$, i.e., 
 $C_1$ = 1.098, $C_2$ = 1.703, $\omega_1 = 2.033 \times 10^{16}$ rad/s, and
$\omega_2 = 1.88 \times 10^{14}$ rad/s \cite{Hough}. The static dielectric constant of SiO$_2$ is thus obtained as $ \epsilon(0) = 3.81$.

For very low disorder variance, the Casimir-vdW interaction dominates and thus the total force between the slabs in vacuum 
 is expected to follow the standard Lifshitz form for neutral dielectrics and thus go from the non-retarded form characterized by 
the power-law decay $\sim D^{-3}$ at small separations, through the retarded form $\sim D^{-4}$ for larger separations 
and then back to the zero-frequency form which for asymptotically large  separations scales again as $\sim D^{-3}$. This behavior
is shown in Fig \ref{fig:fig2}a. Here we change the disorder variance in the range from $g_1=g_2=10^{-11}$ up to $10^{-6}$~nm$^{-3}$ \cite{Kao_Pitaevskii}, showing clearly that the
disorder effects set in for this range of parameters  for separations larger than about 50~nm. This is well into the retarded
regime and is thus beyond the regime where the simple zero-frequency results \cite{jcp2010} can be valid. However, once the disorder effects set in, they 
quickly dominate and the interaction force shows the characteristic power-law decay $\sim D^{-1}$ \cite{prl2010}.  
Note that the magnitude of the total force can increase by orders of magnitude as compared with the 
pure Casimir-vdW force (Fig \ref{fig:fig2}a). Also the transitions between various power-law regimes may depend crucially on 
the characteristics of the dielectric spectra and may be quite complicated for different real materials. 

For identical slabs in vacuum both the Casimir-vdW force as well as the disorder-induced force are attractive. 
When the slabs are dissimilar, one may encounter more interesting cases where the two effects 
oppose each other \cite{jcp2010}. We now consider a situation where
$\epsilon_1(\imath \xi)<\epsilon_m(\imath \xi)<\epsilon_2(\imath \xi)$ and again only the two semi-infinite 
slabs carry disorder charges ($g_m=0$). 
In this case, the pure Casimir-vdW force is known to be repulsive \cite{vdWgeneral,vdw_repulsive,sernelius2} 
but the disorder force may be repulsive or attractive depending
on the static dielectric constants  \cite{jcp2010}. In Fig. \ref{fig:fig2}b, we show the results for a case 
where the disorder force is in fact attractive and thus
for sufficiently large disorder variances in the slabs, one obtains a non-monotonic behavior for the total force as a function of
the separation, $D$. Such a non-monotonic behavior is one of the most remarkable features of the 
interactions between neutral but randomly charged dielectrics.
Non-monotonic fluctuation-induced  interactions between (neutral) materials have 
received a lot of attention in recent years  \cite{metamat,topol,geom_casimir,sernelius}. 
The presence of a small amount of quenched random charges can thus provide another mechanism that can lead to non-monotonic
interactions when dielectric slabs interact across a dielectric medium. 

Note that there is an equilibrium separation distance, $D_0$, where the total force vanishes, which, in the present case, represents a stable equilibrium 
separation between the slabs (Fig. \ref{fig:fig2}b). Thus, the disorder effects give rise to a bound state between neutral slabs which 
otherwise tend to repel each other due to the Casimir-vdW forces. On the other hand, we find a maximum attractive force at slightly larger
separations than $D_0$. This may be used to optimize the thickness of the intervening 
medium in order to achieve the maximum force magnitude between the slabs.
The remarkable point is that, the zero-frequency calculation \cite{jcp2010} (red dashed line in the inset of Fig. \ref{fig:fig2}b) significantly {\em underestimates}
the value of the the bound-state separation; it gives a value of $D_0\simeq 1.4l_{\mathrm{B}}\simeq 79.5$~nm, while the inclusion of higher-order
frequencies (black dashed line in the inset of Fig. \ref{fig:fig2}b)  yields $D_0\simeq 3.4{\ell_\mathrm{B}}\simeq 193.1$~nm for $g_1=g_2=10^{-6}$~nm$^{-2}$
and the parameter values specified in the figure (inset). 
Hence, a systematic calculation of the Casimir-vdW force based on higher-order Matsubara frequencies  
predicts that the above-mentioned non-monotonic behavior can occur in a regime which is even 
more easily accessible to experimental verification \cite{tang}
than the prediction within the zero-frequency calculation \cite{jcp2010}. In the regime that the non-monotonic behavior is found, 
the magnitude of the total force in the presence of disorder is typically much larger than the pure Casimir-vdW force as may be seen 
from Figs. \ref{fig:fig2}b and c.

\begin{figure}[t!]\begin{center}
	\begin{minipage}[b]{0.375\textwidth}\begin{center}
		\includegraphics[width=\textwidth]{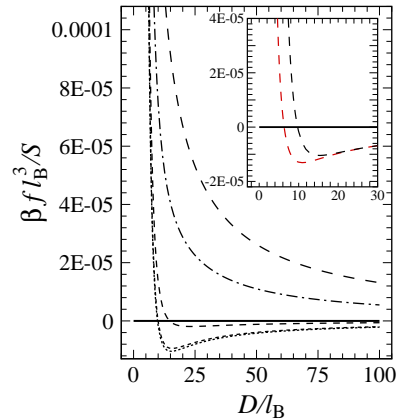} (a)
	\end{center}\end{minipage} \hskip-1cm\vskip2mm
	\begin{minipage}[b]{0.375\textwidth}\begin{center}
		\includegraphics[width=\textwidth]{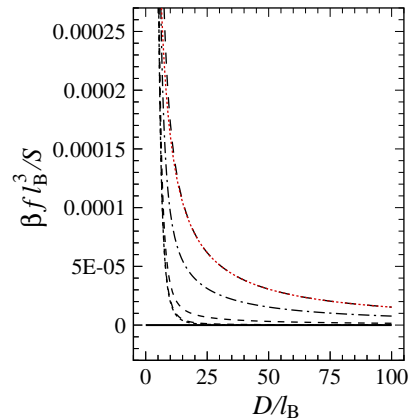} (b)
	\end{center}\end{minipage} 	
\caption{(Color
online)  (a) The rescaled total force, $\beta f l_{\mathrm{B}}^3/S$, 
between two dissimilar net-neutral  slabs when the dielectric response functions satisfy the relationship
$\epsilon_1(\imath \xi)<\epsilon_m(\imath \xi)<\epsilon_2(\imath \xi)$ with the static dielectric constant values
$\varepsilon_1=5$, $\varepsilon_m = 10$ and $\varepsilon_2=50$. Here we assume $g_1=g_2=5\times 10^{-8}\, {\mathrm{nm}}^{-3}$ and 
change $g_m = 10^{-6}, 5\times 10^{-7}, 10^{-7}, 10^{-8}, 10^{-10}\, {\mathrm{nm}}^{-3}$ (from top to 
bottom). The red curve shows the results for $g_m=0$ and  includes only the zero-frequency contribution in the Lifshitz formula. 
Inset shows a closer view of the region around the minimum for this latter case (the point of zero force is shifted 
from $D_0\simeq 6.3l_{\mathrm{B}}\simeq 358$~nm, red curve, to $D_0\simeq 9.7l_{\mathrm{B}}\simeq 551$~nm, black curve).
(b) Same as (a) but for $g_1=g_2=0$. 
}
\label{fig:fig3}
\end{center}
\end{figure}

In the case where the dielectric response functions of the side slabs are smaller than that of the intervening slab 
$\epsilon_1(\imath \xi), \epsilon_2(\imath \xi)<\epsilon_m(\imath \xi)$, one encounters a situation which is opposite to the above case, i.e., 
the pure Casimir-vdW force is attractive \cite{vdWgeneral} but the disorder force is always repulsive and, as expected, 
dominates at sufficiently large separations \cite{jcp2010}, resulting again in 
a non-monotonic behavior for the total interaction force between the slabs with a maximum repulsive force at 
sufficiently high disorder variances as seen in Fig. \ref{fig:fig2}c. 
In this case, the separation distance $D_0$ where the total force vanishes represents an unstable equilibrium distance. 
In other words, the disorder-induced forces in this case not only oppose the Casimir-vdW force but can also give rise to 
a potential barrier in the total interaction free energy. 
Interestingly, in this case the location of the equilibrium moves to larger values of the spacing as well (e.g., from $D_0\simeq 1.7l_{\mathrm{B}}\simeq 96.6$~nm
to $D_0\simeq 2.8l_{\mathrm{B}}\simeq 159.0$~nm for $g_1=g_2=10^{-6}$~nm$^{-2}$
and the parameter values specified in Fig. \ref{fig:fig2}c, inset). This is directly due to the fact that the
inclusion of higher-order Matsubara frequencies leads to a larger attractive Casimir-vdW force.  

\subsection{Role of charge disorder in the intervening medium}
\label{sec:dis}

So far we focused only on cases where the intervening slab does not contain any disorder charges and examined the role of higher-order
Matsubara frequencies. We now proceed by examining the effects that may arise from the presence of charge disorder in the
intervening slab, i.e., $g_m>0$, and compare the results with those we found in the previous Sections and with the zero-frequency results published
elsewhere \cite{jcp2010}, where such effects were not included. 

Let us first consider the case where the dielectric response functions fulfill the relationship 
$\epsilon_1(\imath \xi)<\epsilon_m(\imath \xi)<\epsilon_2(\imath \xi)$. This situation was analyzed in Fig. \ref{fig:fig2}b when there is 
no quenched random charge in intervening slab, i.e., $g_m=0$. We now increase the value of $g_m$ from $10^{-10}$ up to $10^{-6}$~nm$^{-3}$,
while keeping $g_1=g_2 = 5\times 10^{-8}$~nm$^{-3}$ fixed. As seen in Fig. \ref{fig:fig3}a, the non-monotonicity of the interaction profile fades away as
the disorder variance in the intervening slab is increased and the interaction
between the two semi-infinite slabs becomes strongly repulsive.  
This also means that the stable bound-state separation $D_0$ increases and eventually tends to infinity. 
It is easy to see that this situation arises because adding further charges
in the intervening slab is energetically unfavorable, although the slab remains charge neutral on the average. 
In order to demonstrate this effect, we set the disorder variance in the two semi-infinite slabs equal to zero, $g_1=g_2=0$. The total force
in this case is repulsive and increases with $g_m$, Fig. \ref{fig:fig3}b. In fact, one can easily show that for the case with $g_1=g_2=0$,
the disorder-induced force follows  from Eqs. (\ref{free_energy_quenched}) and (\ref{eq:tot_f}) as
\begin{equation}
\frac{\beta f_{\mathrm{dis}}}{S} = \frac{g_m l_{\mathrm{B}}\eta}{2\varepsilon_m D}, 
\label{eq:eta}
\end{equation}
where 
\begin{equation}
\eta = \frac{\varepsilon_1\varepsilon_2-\varepsilon_m^2}{(\varepsilon_1+\varepsilon_m)(\varepsilon_2+\varepsilon_m)}. 
\end{equation}
For the case in Fig. \ref{fig:fig3}b, we have $\eta>0$, and hence the disorder force turns out to be repulsive. In general, when all slabs carry 
disorder charges and assuming that $g_1=g_2=g$, we find
\begin{equation}
\frac{\beta f_{\mathrm{dis}}}{S} = \left[-\frac{g \chi}{(\varepsilon_1+\varepsilon_2)}+\frac{g_m\eta}{\varepsilon_m }\right]\frac{ l_{\mathrm{B}}}{2D}, 
\end{equation}
where 
\begin{equation}
\chi =\frac{(\varepsilon_1 - \varepsilon_m) }{ \varepsilon_2 + \varepsilon_m}
 + \frac{(\varepsilon_2 - \varepsilon_m) }{ \varepsilon_1 + \varepsilon_m}.
\end{equation}
The above formulae clearly show that the disorder-induced force decays quite weakly as $~D^{-1}$. 

\begin{figure}[t!]\begin{center}
	\begin{minipage}[b]{0.375\textwidth}\begin{center}
		\includegraphics[width=\textwidth]{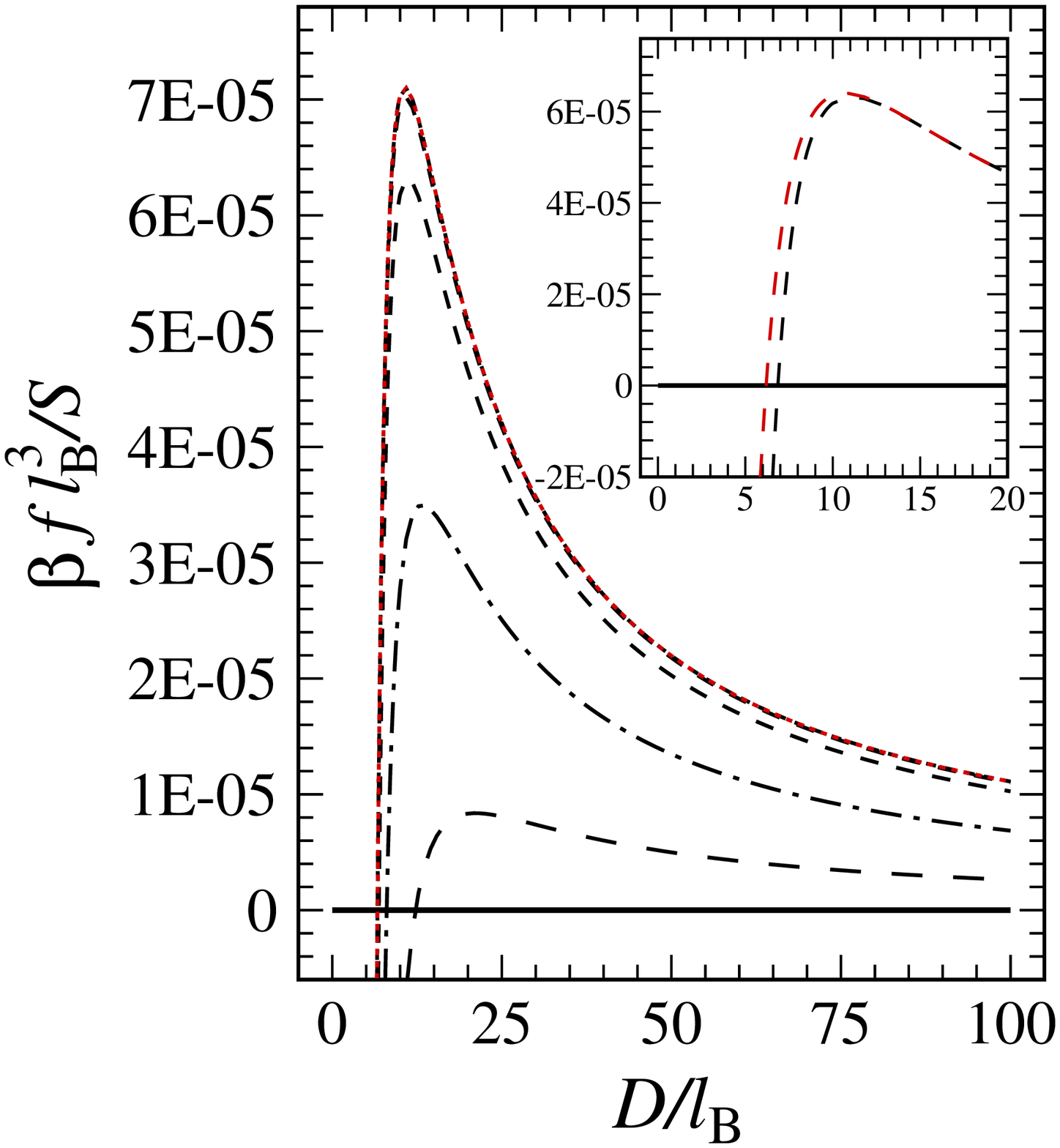} (a)
	\end{center}\end{minipage} \hskip-1cm\vskip2mm
	\begin{minipage}[b]{0.375\textwidth}\begin{center}
		\includegraphics[width=\textwidth]{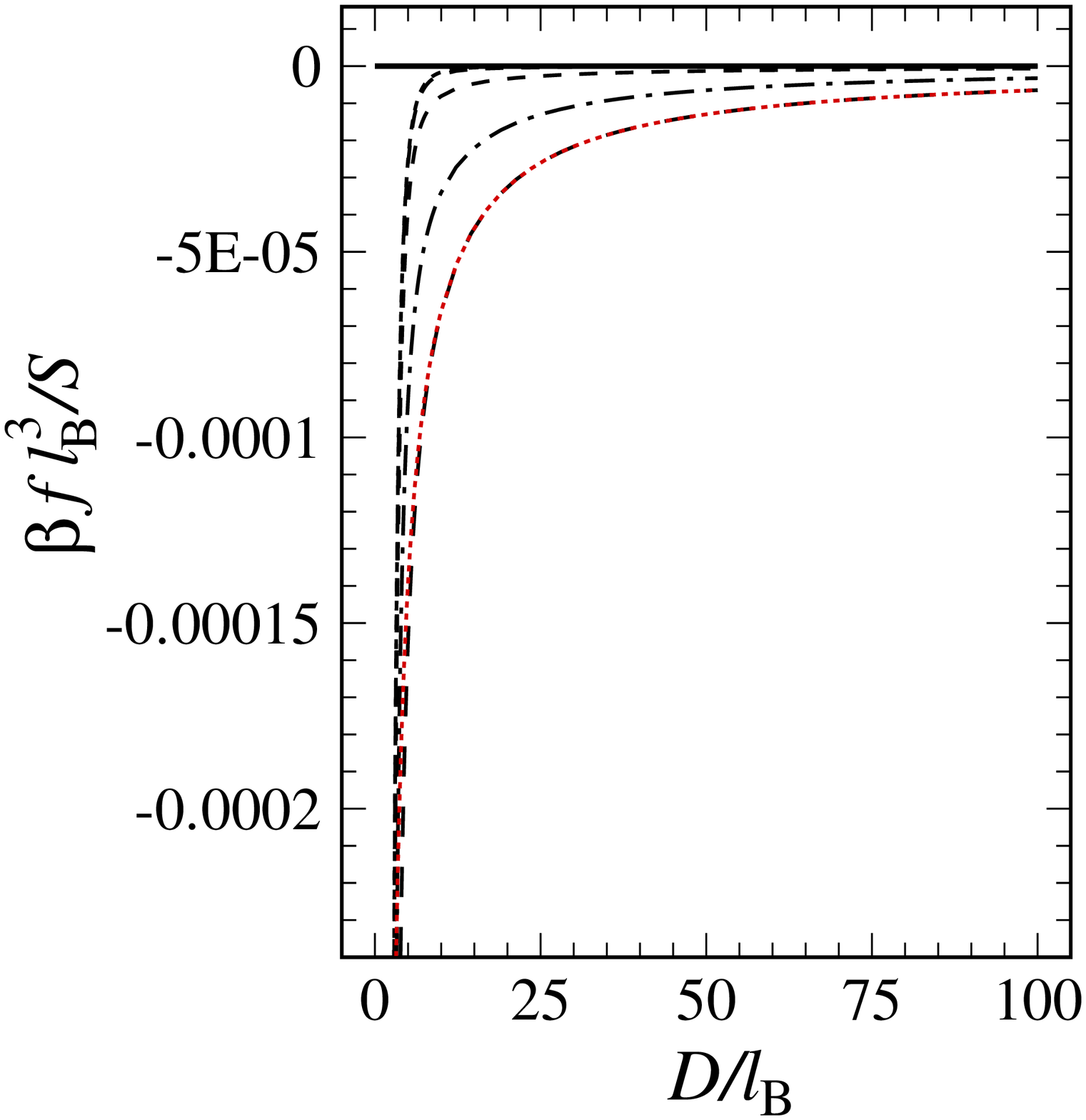} (b)
	\end{center}\end{minipage} 	
\caption{(Color
online)  (a) The rescaled total force, $\beta f l_{\mathrm{B}}^3/S$, between two dissimilar net-neutral slabs when the dielectric response functions satisfy the relationship
$\epsilon_1(\imath \xi), \epsilon_2(\imath \xi)<\epsilon_m(\imath \xi)$ with the static dielectric constant values
$\varepsilon_1=\varepsilon_2=3.81$ (appropriate for SiO$_2$) and $\varepsilon_m = 100$. Here we assume $g_1=g_2=5\times 10^{-8}\, {\mathrm{nm}}^{-3}$ and 
change $g_m = 10^{-6}, 5\times 10^{-7}, 10^{-7}, 10^{-8}, 10^{-10}\, {\mathrm{nm}}^{-3}$ (from bottom to 
top). The red curve shows the results for $g_m=0$ and  includes only the zero-frequency contribution in the Lifshitz formula. 
Inset shows a closer view of the region around the minimum for $g_m = 10^{-7}\, {\mathrm{nm}}^{-3}$ along with the corresponding 
 zero-frequency results (red curve). 
(b) Same as (a) but for $g_1=g_2=0$ and $\varepsilon_1=15$, $\varepsilon_m = 30$ and $\varepsilon_2=25$. 
The red curve shows the  corresponding 
 zero-frequency results for $g_m = 10^{-6}\, {\mathrm{nm}}^{-3}$. 
}
\label{fig:fig4}
\end{center}
\end{figure}

In the case where the dielectric response functions of the bounding slabs are smaller than that of the intervening slabs 
$\epsilon_1(\imath \xi), \epsilon_2(\imath \xi)<\epsilon_m(\imath \xi)$, introducing quenched disorder charges 
in the intervening slab suppresses the maximal repulsive force and shifts the point of zero force $D_0$ to larger
separations as shown in Fig. \ref{fig:fig4}a and the inset. 
Intuitively, this is because the dielectric images for the charges in the intervening slab are of opposite sign
and thus generate attractions when $\varepsilon_1, \varepsilon_2< \varepsilon_m$. 
Equation (\ref{eq:eta}) and the results in Fig. \ref{fig:fig4}b for $g_1=g_2=0$ clearly demonstrates this effect
as we have $\eta<0$ and the only disorder charge contribution comes from those in the intervening slab.  

\begin{figure}[t!]\begin{center}
	\begin{minipage}[b]{0.375\textwidth}\begin{center}
		\includegraphics[width=\textwidth]{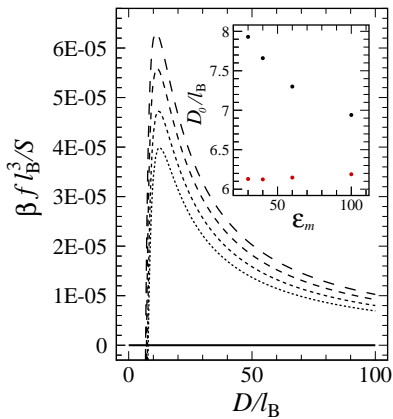} 
	\end{center}\end{minipage}	
\caption{(Color online)  The rescaled total force, $\beta f l_{\mathrm{B}}^3/S$, between two net-neutral  slabs when the dielectric response functions satisfy the relationship
$\epsilon_1(\imath \xi), \epsilon_2(\imath \xi)<\epsilon_m(\imath \xi)$ with the static dielectric constant values
$\varepsilon_1=\varepsilon_2=3.81$ (appropriate for SiO$_2$) and $\varepsilon_m = 30, 40, 60$ and 100 (from bottom to top). 
Here we assume $g_1=g_2=5\times 10^{-8}\, {\mathrm{nm}}^{-3}$ and  $g_m = 10^{-7}\, {\mathrm{nm}}^{-3}$. 
Inset shows the location of the zero-force point, $D_0$, as a function of $\varepsilon_m$ (black dots) compared with the
data when only the zero-frequency contribution is included in the Lifshitz formula (red dots). }
\label{fig:fig5}
\end{center}
\end{figure}

The above results strongly depend on the dielectric difference between the slabs. In fact, the disorder-induced force, Eq. (\ref{eq:eta}), shows
a non-monotonic dependence on $\varepsilon_m$. If we assume $\varepsilon_1=\varepsilon_2=\varepsilon<\varepsilon_m$, the attractive disorder force will
increases in magnitude with $\varepsilon_m$ when $\varepsilon_m<(1+\sqrt{2})\varepsilon$ and deceases in magnitude otherwise. 
For instance, in Fig. \ref{fig:fig5}, we choose two identical slabs with $\varepsilon_1=\varepsilon_2=3.81$ but change the 
static dielectric constant of the intervening slab from $\varepsilon_m=30$ up to 100. The total force becomes increasingly more repulsive and the distance
at which the force becomes zero, $D_0$, decreases significantly as $\varepsilon_m$ is increased as shown in the inset of Fig. \ref{fig:fig5} (black dots). 
It should be noted however that the zero-force separation is much larger than
what is expected from a zero-frequency calculation \cite{jcp2010}, which varies only weakly with $\varepsilon_m$ (inset, red dots). 

\section{Conclusion}
\label{sec:discussion}

We have studied interactions between randomly charged but otherwise net-neutral slabs by including first, the full spectrum of 
electromagnetic field fluctuations and second, by taking into account  the presence of random charges in the medium between the two slabs. This
is done by employing the Lifshitz theory which includes a summation over all Matsubara
frequencies, leading to a direct generalization of our previous results \cite{prl2010,jcp2010}, which were 
obtained based only on the zero-frequency effects and were thus valid at large separations. 
This is a crucial new aspect of our analysis of charge disorder as it extends the regime of validity of some of the 
key findings to the full range of inter-slab separations, especially, down to the nano scale (as long as the continuum 
model assumed within the Lifshitz theory remains valid).

We can thus draw certain important conclusions regarding the role of disorder effects. In particular, it is shown that the characteristic 
$\sim D^{-1}$ behavior due to quenched disorder \cite{prl2010} sets in well within the retarded regime (e.g., around 50-500~nm in Fig. \ref{fig:fig2}a)
and thus the force curves deviate
rapidly from the standard (retarded) $\sim D^{-4}$ Casimir-vdW behavior even for highly clean samples with disorder variances down to
$10^{-9}\, {\mathrm{nm}}^{-3}$. This behavior should be contrasted with the zero-frequency theory  \cite{jcp2010} which 
predicts that the $\sim D^{-1}$ emerges due to a crossover from the classical  $\sim D^{-3}$ behavior. 

Another remarkable prediction that follows from our present analysis is that the non-monotonic behavior 
of the total force as a function of distance persists when higher-order Matsubara frequencies are included; 
however, in almost all cases, the (stable or unstable) equilibrium separation is shifted to much larger values
than  predicted  based solely on the zero-frequency theory \cite{jcp2010}. 
This is important in that it predicts that the disorder-generated non-monotonic behavior of the force is more easily amenable to experimental measurements
that expected from zero-frequency calculations. 

We have also shown that the presence of charge disorder in the intervening medium  leads to another additive contribution to the total force that can be repulsive or attractive depending
on the system parameters (such as the relationship between dielectric response functions of the media) and can, for instance,
wash out the non-monotonic behavior of the total force. Such non-monotonic behaviors for the interaction 
between dielectric slabs have received a lot of attention in the context of the Casimir effect and are 
known to emerge, e.g., in the case of metamaterials \cite{metamat} and/or other exotic materials such as topological 
insulators  \cite{topol}, as well as in certain non-trivial geometries \cite{geom_casimir,sernelius}.
In our analysis the  $\sim D^{-1}$  behavior of the force for identical slabs and the  non-monotonic force
profile for dissimilar slabs represent characteristic fingerprints of the charge disorder and can thus be useful in assessing 
whether the experimentally observed interactions in ordinary dielectrics can be interpreted  in terms of disorder effects.  

We presented our results explicitly for the case where the disorder distribution is statistically homogeneous and uncorrelated in space, although the  formalism
 in Section \ref{sec:model} is  in general applicable to layered materials exhibiting finite lateral  correlations
for the disorder charges as well. Thus, it would be interesting in the future to examine the above-mentioned effects in the situation where
the disorder distribution consists of random patches (domains) of finite size \cite{jcp2010}, and specifically, when the layering assumption 
is relaxed and the correlation domains within the slabs have a three-dimensional structure.

\section{Acknowledgments}

A.N. is supported by a Newton International Fellowship from the Royal Society, the Royal Academy of Engineering, 
and the British Academy. 
 R.P. acknowledges support from ARRS through the program P1-0055 and the research project J1-0908. 


\end{document}